\documentclass[prl,twocolumn,floatfix]{revtex4}
\pdfoutput=1
\usepackage{amsmath,amssymb,graphicx,bm,url}

\newcommand{\avr}[1]{\left \langle #1 \right \rangle}

\begin{document}

\title{Neutrality point of graphene with coplanar charged impurities}

\author{Michael M. Fogler}

\affiliation{Department of Physics, University of California San Diego, La
Jolla, 9500 Gilman Drive, California 92093}

\date{v1: October 16, 2008, v2: November 12, 2009}

\begin{abstract}

The ground-state and the transport properties of graphene subject to the
potential of in-plane charged impurities are studied. The screening of the impurity
potential is shown to be nonlinear, producing a fractal structure of electron and hole puddles. Statistical properties of this density distribution as well as the
charge compressibility of the system are calculated in the leading-log
approximation. The conductivity depends logarithmically on $\alpha$,
the dimensionless strength of the Coulomb interaction. The theory is
asymptotically exact when $\alpha$ is small, which is the case for graphene on a
substrate with a high dielectric constant.

\end{abstract}


\maketitle


A number of recent experimental~\cite{Geim2007rg, Tan2007mos, Chen2008cis,
Jang2008tte, Martin2008oeh} and theoretical~\cite{Ando2006sea, Katsnelson2006nso, Nomura2007qtm, Shytov2007vpa, Fogler2007soa, Pereira2007cip,
Ostrovsky2006eti, Adam2007sct, Shklovskii2007smo, Rossi2008gsg, Rossi2008emt}
investigations have studied the effect of \emph{charged} impurities on the
properties of graphene~\cite{Castroneto2009tep} at its neutrality point (NP). It has been shown that in response to the random potential of impurities electron density becomes very inhomogeneous. A deeper physical understanding of these inhomogeneities is important for clarifying the true nature of disorder in this new material. Hence, an analytic approach to this problem is desirable. Below I present my results in this direction.

As a model I assume that impurities with random charge $\pm e$ and total
concentration $n_i$ are distributed randomly on a graphene sheet of area $A$.
The system resides in a medium of dielectric constant $\kappa$. The
strength of the Coulomb interaction, $U(r) = e^2 / \kappa r$, is characterized
by the parameter
$
\alpha = {e^2}/\,{\kappa \hbar v}
$
where $v$ is the Fermi velocity.

My main results are as follows.
For $\kappa \gg 1$ where the interaction is weak, $\alpha \ll 1$,
the basic electronic properties of graphene at the NP can be computed to the
leading order in $1 / \mathcal{L}$, where $\mathcal{L} \gg 1$ is the solution of the equation
\begin{equation} \label{eqn:L}
\mathcal{L} = \ln \big(1 / (4 \alpha \mathcal{L}) \big)\,.
\end{equation}
I find that the charge
compressibility $\chi_0$ at the NP (measured in Ref.~\onlinecite{Martin2008oeh}) is given by 
\begin{equation} \label{eqn:chi_0}
\chi_0 = {\kappa} / (2 \pi e^2 R)\,,
\quad 
             R = {\ell} / ({4 \alpha \mathcal{L}})\,,
\end{equation}
where $R$ is the screening length and $\ell = ({ 2 \alpha \sqrt{n_i} }\,\,)^{-1} \ll R$ has the meaning of a typical quantum uncertainty in the
quasiparticle positions. Observables, such as the
correlator $S(r) \equiv \avr{n(\mathbf{r}) n(0)}$ of the local
density become nontrivial on scales $r >
\ell$. For such $r$, I find
\begin{align}
S(r) &= \frac{\mathcal{L}^2}{2 \pi \ell^4} \left[
     3 \vartheta \sqrt{1 - \vartheta^2}
     + \left(1 + 2 \vartheta^2 \right) \arcsin \vartheta \right]
\,,
\label{eqn:S_r}\\
\vartheta(r) &= \frac{K(r)}{K(\ell)}\,,
\quad
K(\ell) = \frac{\pi}{2} \left( \frac{\hbar v}{\ell} \right)^2 \mathcal{L}\,.
\label{eqn:vartheta}
\end{align}
Here $K(r) \equiv \avr{\Phi(0) \Phi(\mathbf{r})}$ is the correlator
of the screened random potential $\Phi$. I show that $K(r)$ behaves as
\begin{equation}
K(r) = \frac{\pi}{2} \left( \frac{\hbar v}{\ell} \right)^2 \times
\biggl\{
\begin{alignedat}{2}
&\ln ({R}/{r})\,,&  &\quad \ell \ll r \ll R\,,
\\
&2 (R / r)^3,    &  &\quad R \ll r\,.
\end{alignedat}
\label{eqn:K_r}
\end{equation}
Equation~\eqref{eqn:K_r} agrees with the results
of Refs.~\onlinecite{Adam2007sct} and \onlinecite{Galitski2007srv} in
the limit $\alpha \ll 1$. Equation~\eqref{eqn:S_r} is in a qualitative
agreement with the numerical simulations~\cite{Rossi2008gsg} for $\alpha
\sim 1$ (where my theory is at the border of validity).

Of special interest for transport and imaging experiments are the contours
$n(\mathbf{r}) = 0$, which separate electron [$n(\mathbf{r}) > 0$] and hole
[$n(\mathbf{r}) < 0$] regions. Only contours larger than some minimum size $\Lambda^{-1}$ are important for the key observables (in imaging, $\Lambda^{-1}$ is the spatial resolution; in transport, $\Lambda^{-1} \sim \ell \sqrt{\mathcal{L}}$, see below). To isolate them, we can separate the total density and the total potential into smoothly and rapidly varying parts,
\begin{equation} \label{eqn:n_smooth}
n(\mathbf{r}) = \bar{n}(\mathbf{r}) + \delta n(\mathbf{r})\,,
\quad
\Phi(\mathbf{r}) = \bar{\Phi}(\mathbf{r}) + \delta \Phi(\mathbf{r})\,,
\end{equation}
where $\bar{n}(\mathbf{r})$ and $\bar{\Phi}(\mathbf{r})$ contain only Fourier harmonics with $k < \Lambda$. Below I assume that
$1 / \ell \lesssim \Lambda \ll 1 / R$. I refer to the contours $\bar{n}(\mathbf{r}) = 0$ as the $p$-$n$ junctions (PNJ). I show that most of the PNJ are closed loops of diameter $d \sim \Lambda^{-1}$. They reside inside of successively larger loops, forming a self-similar set. The fractal dimension $D_h$ of the PNJ contours is equal to $3/2$ at $d < R$ but
becomes $7/4$ at $d > R$.
Finally, the conductivity at the NP is
\begin{equation} \label{eqn:sigma_NP}
\sigma_\text{NP} = ({e^2} / {h}) c \mathcal{L}\,,
\quad
c = 0.50 \pm 0.05\,.
\end{equation}
This result is supposed to be valid at temperatures high enough that
weak localization effects~\cite{Foster2008gvl} can be neglected but low
enough that elastic scattering by impurities is still the dominant
current relaxation mechanism.

A similar expression for $\sigma_\text{NP}$ can be deduced from
Ref.~\onlinecite{Adam2007sct}. It has $c = 4 / \pi$ for $\alpha \ll 1$,
which is larger than my value of $c$ by about a factor of two.
Numerical estimates of $\sigma_\text{NP}$ at $\alpha \sim 1$ were
reported in Ref.~\onlinecite{Rossi2008emt}.

Let us turn to the derivation.
Let $V(\mathbf{r})$ be the bare random potential due to impurities. In the
$k$-space it is given by ${V}_{\mathbf{k}} = U_\mathbf{k} {N}_{\mathbf{k}}$, where ${N}_{\mathbf{k}}$ and $U_{\mathbf{k}} = 2 \pi \alpha \hbar v / k$ are the Fourier transforms of
the impurity density and the Coulomb potential, respectively. At $0 < k \ll n_i^{1/2}$ we can treat them as zero-mean Gaussian random variables with variances
$\avr{{N}_{\mathbf{k}} {N}_{\mathbf{k}^\prime}}
= A n_i \delta_{\mathbf{k}, {-\mathbf{k}^\prime}}$ and
$\avr{{V}_{\mathbf{k}} {V}_{\mathbf{k}^\prime}}
= U_{\mathbf{k}}^2 \avr{{N}_{\mathbf{k}} {N}_{\mathbf{k}^\prime}}$. It is easy to see then that the coarse-grained random potential $\bar{V}(\mathbf{r})$ is infrared divergent~\cite{Efros1993dos}:
\begin{equation}
\avr{\bar{V}^2(\mathbf{r})} = \sum_{k < \Lambda} \avr{{V}_{\mathbf{k}} {V}_{\mathbf{k}^\prime}}
\simeq  ({\pi} / {2}) ( {\hbar v} / {\ell} )^2
 \ln(\Lambda \sqrt{A}\,)\,.
\label{eqn:VV}
\end{equation}
This divergence is of course cured once the bare potential is replaced by a screened one,
\begin{equation} \label{eqn:Phi_k}
\Phi_\mathbf{k} = V_\mathbf{k} + n_\mathbf{k} U_\mathbf{k}
= V_\mathbf{k} + (2 \pi \alpha \hbar v / k) n_\mathbf{k}\,,
\end{equation}
which is expected to have the following properties:
\begin{equation}
\Phi_{\mathbf{k}} = V_{\mathbf{k}}\,, \:\:\: k R \gg 1\,
\text{;\:\:\:}
\Phi_{\mathbf{k}} = V_{\mathbf{k}} \frac{k R}{1 + k R}\,, \:\:\: k R \ll 1\,.
\label{eqn:Phi_vs_V}
\end{equation}
The last equation follows from the definition of the thermodynamic charge
compressibility $\chi_0 = -n_{\mathbf{k}} / \Phi_{\mathbf{k}}$
and the relation between $\chi_0$ and $R$, Eq.~\eqref{eqn:chi_0}.
I wish to show that $R$ is given by the second formula in Eq.~\eqref{eqn:chi_0}, in particular, that it is a function of the impurity concentration. This means that the screening is nonlinear, i.e., the response of electrons in graphene to long-wavelength Fourier harmonics $k < R^{-1}$ is non-perturbative in $V_{\mathbf{k}}$.

Formally, $n(\mathbf{r})$ and $\Phi(\mathbf{r})$ can be sought by minimizing the total energy $E[n] + \mathcal{E}[n]$, where
$
E[n] = A^{-1}\, \sum_{\mathbf{k}} 
      {V}_{-\mathbf{k}} {n}_{\mathbf{k}}
    + (1 / 2) {U}_{\mathbf{k}} |{n}_{\mathbf{k}}|^2
$
is the electrostatic energy and $\mathcal{E}[n]$ is the sum of kinetic, exchange, and correlation energies~\cite{Hohenberg1964}. Since the functional $\mathcal{E}[n]$ is nonlocal, nonlinear, and strictly speaking, unknown, further steps require approximations. We are aided by the parameter $\mathcal{L}$: the condition $\mathcal{L} \gg 1$ ensures separation of two length scales: (i) the typical quantum uncertainly $\ell$ over which the nonlocality of $\mathcal{E}[n]$ is important and (ii) the screening length $R \gg \ell$ at which the nonlinear screening sets in. As a result, a \emph{local} density approximation $\mathcal{E} = \int\! d \mathbf{r}\, \varepsilon \big( \bar{n}(\mathbf{r}) \big)$ is valid and the total energy is minimized when
\begin{equation}\label{eqn:MTFA_general}
\bar{\Phi}(\mathbf{r})
 = -\varepsilon^\prime\big(\bar{n}(\mathbf{r})\big)\,.
\end{equation}
The kernel $\varepsilon(n)$ can be computed by integrating out the rapid density fluctuations $\delta n(\mathbf{r})$ within the linear-response theory~\cite{Ando2006sea, Gorbar2002mfd} for $\alpha \ll 1$. The result is~\cite{EPAPS}
\begin{equation}\label{eqn:g_near}
\varepsilon = \frac{2 \sqrt{\pi}}{3}\, \hbar v |\bar{n}|^{3/2}
                 + \delta \varepsilon\,,
\quad
\delta \varepsilon = -\frac{3 \varepsilon}{4 |\bar{n}| \ell^2}
 \ln \left(\frac{k_F}{\Lambda} \right).
\end{equation}
Here $k_F(\mathbf{r}) = \sqrt{ \pi \raisebox{0pt}[1.75ex]{$|\bar{n}(\mathbf{r})|$}}\,$ is the local Fermi momentum. Equation~\eqref{eqn:MTFA_general} can now be easily inverted to give
\begin{equation} \label{eqn:MTFA}
\bar{n}(\bar{\Phi}) \simeq
               -\frac{\bar{\Phi} \left|\bar{\Phi} \right|} {\pi (\hbar v)^2}
               - \frac{ \text{sgn}\left(\bar{\Phi}\right)}{2 \ell^2}\,
                \ln  \frac{\left| \bar{\Phi} \right|}{\hbar v \Lambda}
                \,,
\quad
\left| \bar{\Phi} \right| \gg \hbar v \Lambda\,.
\end{equation}
Later, we will also need the derivative of function $\bar{n}(\bar{\Phi})$:
\begin{equation} \label{eqn:chi_r}
\chi(\bar{\Phi}) \equiv -\frac{d \bar{n}}{d \bar{\Phi}}
 \simeq \frac{2}{\pi} \frac{\left|\bar{\Phi}\right|}{(\hbar v)^2}
              + \frac{1}{2 \ell^2 \left|\bar{\Phi}\right|}\,,
\quad
\left| \bar{\Phi} \right| \gg \hbar v \Lambda\,.
\end{equation}
We have reduced the original \emph{quantum} problem to the nonlinear integral equation \eqref{eqn:MTFA} for the \emph{classical} quantity $\bar{n}(\mathbf{r})$ [Note that $\bar{\Phi}[\bar{n}]$ is given by Eq.~\eqref{eqn:Phi_k}.] I call this a renormalized Thomas-Fermi approximation (RTFA)~\cite{Comment_on_MTFA}.

In conventional gapped semiconductors, this would be about as far as one could go analytically before having to resort to heuristic estimates~\cite{Efros1984}, computer simulations~\cite{Efros1993dos}, or variational methods~\cite{Fogler2004nsa}. The reason is as
follows: the energy minimization is constrained by $\bar{n}(\mathbf{r}) \geq 0$, leading to a finite area fraction of depletion regions $\bar{n}(\mathbf{r}) = 0$. Therein Eq.~\eqref{eqn:MTFA_general} is replaced by $\Phi(\mathbf{r}) \geq 0$~\cite{Efros1984}, so that no analog of Eq.~\eqref{eqn:MTFA} exists, making the problem analytically intractable. In graphene, there is no constraint on the sign of density. Instead, the conditions for validity
of Eq.~\eqref{eqn:g_near} are $\delta \varepsilon \ll \varepsilon$ and $\Lambda \ll k_F$, which are typically satisfied everywhere except near the PNJ (see below). However, this kind of ``depletion regions'' occupy a parametrically small area fraction.

Yet another serendipity is that the divergence of the bare potential is only logarithmic, cf.~Eq.~\eqref{eqn:VV}. The enables us to compute the non-perturbative long-range response with a logarithmic accuracy. Indeed, in view of Eq.~\eqref{eqn:Phi_vs_V}, $\Phi_\mathbf{k}$ and $V_\mathbf{k}$ are linearly related at all $\mathbf{k}$, except perhaps $k \sim 1 / R$. But this ``difficult'' range of intermediate $k$ makes an $\mathcal{O}(\mathcal{L}^{-1})$ contribution to the total $\bar{\Phi}(\mathbf{r})$. To this order in $\mathcal{L}$ we can treat $\bar{\Phi}(\mathbf{r})$ as a Gaussian random potential. Its probability distribution function (PDF) is uniquely determined by the correlator $\bar{K}(r) \equiv \avr{\bar{\Phi}(0) \bar{\Phi}(\mathbf{r}) }$, to calculate which we \emph{can} use Eq.~\eqref{eqn:Phi_vs_V}. Since $\bar{K}(r) \simeq K(r)$ for $r \Lambda \gg 1$, this immediately leads to Eq.~\eqref{eqn:K_r}.

Next, Gaussian statistics implies ergodicity. Therefore, the density correlation function can be written as
\begin{equation}
S(r) = \int d f d f^\prime\, P_{\Phi \Phi}(r, f, f^\prime\,)
            \bar{n}(f)\, \bar{n}(f^\prime\,)\,,
\label{eqn:S}
\end{equation}
where $P_{\Phi \Phi} \equiv \big\langle \delta \big(f -
\bar{\Phi}(0)\big)\, \delta \big(f^\prime - \bar{\Phi}(\mathbf{r})\big)
\big\rangle$ is the two-point PDF of $\bar{\Phi}$ and function $\bar{n}(f)$ is given by Eq.~\eqref{eqn:MTFA}. Using the standard expression~\cite{Isichenko1992pst} for $P_{\Phi \Phi}$ in terms of $\bar{K}$ and a bit of algebra~\cite{EPAPS}, I get Eq.~\eqref{eqn:S_r}. At $r \gg R$,
it simplifies to
$S(r) \simeq \langle \chi \rangle^2 K(r)$, where
$\langle \chi \rangle \equiv \int d f P_\Phi(f)
 \chi(f)$,
which implies that $\chi_0 = \langle \chi \rangle$. Substituting
here Eq.~\eqref{eqn:chi_r} and computing the average over the Gaussian field $\bar{\Phi}$ using Eq.~\eqref{eqn:K_r}, I get
$
\chi_0^2 = {8} {K(\ell)} / (\pi^3 \hbar^4 v^4)
$.
Finally, since $K(\ell)$ is given by Eq.~\eqref{eqn:vartheta}, we recover the desired Eq.~\eqref{eqn:chi_0}.

It is also possible to compute the PDF $P_n$ of $\bar{n}$. The Gaussian statistics of $\bar{\Phi}$ combined with Eq.~\eqref{eqn:MTFA} entails
\begin{equation} \label{eqn:P_n}
P_n(\bar{n})
= \frac{\ell}{2 \sqrt{\pi \mathcal{L} |\bar{n}|}}
\,
 \exp\left(- \frac{|\bar{n}| \ell^2}{\mathcal{L}} \right)
\,, \quad |\bar{n}| \ell^2 \gg 1
\,.
\end{equation}
Note, however, that this equation is invalid at small densities, $|\bar{n}| < 1 / \ell^2$, where the RTFA fails. At such $\bar{n}$
the divergence of $P_n(\bar{n})$ basically saturates~\cite{EPAPS}.
These low-density regions are usually found near the PNJ.

Geometrically, the PNJ are the isolines $\bar{\Phi}(\mathbf{r}) = 0$ of a surface with the ``height'' profile $\bar{\Phi}({\mathbf r})$. The zero height is the percolation threshold, and so all but one of the PNJ are closed loops. These loops are characterized by a certain fractal dimension $D_h$. For loops of diameter $d$ in the range $\Lambda^{-1} \ll d \ll R$, in which
$\bar{\Phi}(\mathbf{r})$ surface is logarithmically rough
[Eq.~\eqref{eqn:K_r}], we have the exact
result~\cite{Kondev2000nmf} $D_h = 3 / 2$, which means that closed-loop PNJ
typically have the perimeter length of
\begin{equation} \label{eqn:p}
            p \sim d^{3/2} \ell^{-1 / 2}\,,
\quad
            \Lambda^{-1} \ll d \ll R\,.
\end{equation}
As $d$ increases beyond $R$, the correlator $K(d)$ rapidly decays [Eq.~\eqref{eqn:K_r}], and so $D_h$ crosses over~\cite{Kondev2000nmf} to the usual uncorrelated percolation
exponent~\cite{Isichenko1992pst} of $7 / 4$.

Let us now discuss electron transport.
Away from the NP where electron density $n$ is large and
homogeneous, transport can be studied by means of the usual kinetic
equation. The final results for conductivity $\sigma(n)$
and transport mean-free path $l(n)$ are~\cite{Ando2006sea,
Nomura2007qtm, Ostrovsky2006eti, Adam2007sct}
\begin{equation}
\sigma(n) = \frac{8}{\pi}\, \frac{e^2}{h}\, \ell^2 |n|\,,
\quad\
   l(n) = \frac{4}{\pi^{3/2}}\, \ell^2 |n|^{1/2}.
\label{eqn:l_tr}
\end{equation}
Our goal is to compute the conductivity $\sigma_\text{NP}$ at the NP where
$n = n(\mathbf{r})$ is inhomogeneous.

The first step is to show that we can define the conductivity
$\bar{\sigma}(\mathbf{r})$ and the mean-free path $\bar{l}(\mathbf{r})$ locally.
Under the assumed condition $1 / R \ll \Lambda \lesssim 1 / \ell$ for typical
$\bar{n}(\mathbf{r}) \sim \mathcal{L} / \ell^2$ we have $\nabla l(\bar{n}) \sim
\Lambda l(\bar{n}) \lesssim 1$. Hence, function $l\big(\bar{n}(\mathbf{r})\big)$
is slowly varying and Eq.~\eqref{eqn:l_tr} for the uniform density can be used:
$\bar{l} = l(\bar{n})$. In turn, the local conductivity is given by the Einstein
relation $\bar{\sigma}(\mathbf{r}) = e^2 \chi(\mathbf{r}) v \bar{l}(\mathbf{r})
/ 2$. To be careful one should check this zeroth order result by calculating
higher order corrections due to spatial fluctuations of the collision term in
the kinetic equation. This can be done by treating $\delta l^{-1}(\mathbf{r})
\equiv l^{-1}\big(n(\mathbf{r})\big) - l^{-1}(\bar{n})$ as a perturbation.
Skipping the details~\cite{EPAPS}, I just announce the conclusion: if $|\bar{n}| \gg 1 /
\ell^2$, then the fluctuations of $\delta l^{-1}(\mathbf{r})$ are self-averaging
on the scale of $l(\bar{n})$. In other words, corrections to $l(\bar{n})$ are parametrically small. For simplicity, I
will ignore them together with the similar corrections to $\chi(\mathbf{r})$
[the second term in Eq.~\eqref{eqn:chi_r}] to obtain
\begin{equation} \label{eqn:sigma_large}
  \bar{\sigma}(\mathbf{r}) \simeq \sigma\big(\bar{n}(\mathbf{r})\big)\,,
\quad
 |\bar{n}(\mathbf{r})| \gg 1 / \ell^2.
\end{equation}
Problems arise at $|\bar{n}| \lesssim 1 / \ell^2$ where the corrections to
Eq.~\eqref{eqn:sigma_large} exceed $\sigma(\bar{n})$. In the classical regime
$\bar{\sigma} \gg e^2 / h$, we could have handled this by adopting
a model form
\begin{equation} \label{eqn:sigma_PNJ}
\bar{\sigma}(\mathbf{r}) = \sigma_0 \zeta(\mathbf{r})\,, \quad
|\bar{n}| \lesssim 1 / \ell^2,
\end{equation}
where $\zeta(\mathbf{r}) > 0$ is some random function with the correlation length
$\Lambda^{-1} \sim \ell$ and the typical value of the order of unity~\cite{Comment_on_ballistic}. However, in graphene $\sigma_0 \equiv \sigma(1 / \ell^2) = (8 / \pi) e^2 / h$ is so low that the very concept of local conductivity is potentially jeopardized by quantum interference and localization effects. Fortunately, Eq.~\eqref{eqn:sigma_PNJ} is saved by the special geometry of the regions where it is intended to be used. These regions typically form ribbons of width $x \sim \ell$ that follow
the PNJ loops. These low-conductance ``ribbons'' are connected to
high-conductance ``reservoirs'' on both sides of the PNJ. In such a geometry the
importance of the quantum effects is controlled~\cite{Beenakker1997rmt} by the
total conductance $G_p$ across the perimeter length $p$ of the PNJ.
For $p \gg \ell$, it is given by
\begin{equation} \label{eqn:sigma_p}
G_p \sim \sigma_0\, \frac{p}{\ell} \gg \frac{e^2}{h}\,,
\end{equation}
in which case the semiclassical model~\eqref{eqn:sigma_PNJ} is
justified. Again we have succeeded in reducing the complicated quantum problem to a simpler classical one: finding the macroscopic
conductivity $\sigma_\text{NP}$ of an inhomogeneous medium with local conductivity given by Eqs.~\eqref{eqn:sigma_large} and \eqref{eqn:sigma_PNJ} as a function of the local density $\bar{n}(\mathbf{r})$.

A rigorous upper bound on $\sigma_\text{NP}$ is the spatially averaged conductivity. Due to ergodicity of Gaussian fields, the averaging can be done over $\bar{n}$ instead. Using Eqs.~\eqref{eqn:P_n} and \eqref{eqn:l_tr}, I obtain:
\begin{equation} \label{eqn:sigma_bounds}
\sigma_\text{NP} < \langle \bar{\sigma} \rangle = \int
d \bar{n} P_n(\bar{n}) \bar{\sigma}(\bar{n})
 \simeq (4 \mathcal{L} / \pi) e^2 / h\,.
\end{equation}
(Incidentally, this enables me to conclude that Ref.~\onlinecite{Adam2007sct}
overestimates $\sigma_\text{NP}$ in the limit $\alpha \ll 1$.)

Now I present the argument crucial to my theory of transport. It shows
that $\sigma_\text{NP}$ is \emph{not} sensitive to the details of the
model form~\eqref{eqn:sigma_PNJ}. Instead, it is determined by the typical local
conductivity, and so is not far below the upper bound~\eqref{eqn:sigma_bounds}.
This statement should be contrasted with other theoretical views on the subject. Since the low-conductivity regions~\eqref{eqn:sigma_PNJ} reside at the percolation contour, it has been suggested~\cite{Geim2007rg, Adam2007sct, Cheianov2007rrn}
that the transport at the NP may be governed by
percolation~\cite{Isichenko1992pst}. In that picture the current paths are
severely constrained in order to avoid crossing the PNJ as much as possible and
$\sigma_\text{NP}$ depends on their average
transparency~\cite{Cheianov2007rrn}.

I show that in the model under study the closed loop PNJ contours are \emph{not} resistive
enough for the percolation effects to develop. Indeed, the percolation
approach~\cite{Cheianov2007rrn} would apply only if there existed a wide range
of loop diameters $d \gg \ell$ such that the conductance $G_p$ of the loop perimeter
were much lower than the conductance $G_d$ of their interior.
Suppose first that $d \ll R$ for such loops. Let us show that it leads
to a contradiction. If the current avoids crossing the perimeter, it has to
flow in narrow channels of some width $w \ll d$ due to the fractal geometry of
the loop. Hence, $G_d \sim \bar{\sigma}_\text{int} w / d \ll
\sigma_\text{int}$ where $\sigma_\text{int}$ is the typical local conductivity
inside the loop. To estimate the latter, I note that from the general properties of Gaussian random fields~\cite{Isichenko1992pst} the potential $\bar{\Phi}$ at a distance $r$ from the PNJ ($\bar{\Phi} = 0$) has the variance
\begin{equation}
\avr{\bar{\Phi}^2}_{\text{PNJ}} =
 K(\ell) - \frac{K^2(r)}{K(\ell)} \simeq
\frac{\pi}{4} \left( \frac{\hbar v}{\ell} \right)^2
\ln \left(\frac{r}{\ell}\right),
\label{eqn:Phi_PNJ}
\end{equation}
so that the corresponding density is $\bar{n} \sim \ln (r /
\ell) /  \ell^2$. Using $r \sim d$ and Eq.~\eqref{eqn:l_tr}, I get $\sigma_\text{int} \sim \sigma_0 \ln (d / \ell)$.
In comparison, $G_p \sim \sigma_0 (d / \ell)^{3 / 2}$, cf.~Eqs.~\eqref{eqn:p} and \eqref{eqn:sigma_p}. Hence, $G_p \gg G_d$, which contradicts the assumption made
(the same is true for $d \gg R$). I conclude that
in the present model the percolation-type transport is not realized.

In the absence of strong ramification of current paths by the PNJ, I expect the numerical coefficient $c$ in Eq.~\eqref{eqn:sigma_NP} to be of the order of unity. To estimate it more accurately I use the effective medium theory (EMT). Indeed, the EMT is usually adequate for systems where percolation effects are unimportant.
Within the EMT, $\sigma_\text{NP}$ is determined from a certian nonlinear
equation, the two most popular versions of which were originally proposed in
Ref.~\onlinecite{Bruggeman1935} and Ref.~\onlinecite{Hori1975}:
\begin{alignat}{2}
 \avr{ \frac{\sigma_\text{NP}}{\bar{\sigma} + (\mathcal{D} - 1) \sigma_\text{NP}} }
 &= \frac{1}{\mathcal{D}} & &\quad \text{(Bruggeman)}\,,
\label{eqn:Bruggeman}\\
\int\limits_0^\infty \frac{d z}{e^{z}} \ln \avr{ 
\exp \left[ \frac{z}{\mathcal{D}} \frac{\bar{\sigma}}{\sigma_\text{NP}} \right] }
 &= \frac{1}{\mathcal{D}} & &\quad \text{(Hori)}\,.
\label{eqn:Hori}
\end{alignat}
Here $\mathcal{D} = 2$ is the space dimension. Equations~\eqref{eqn:Bruggeman} and
\eqref{eqn:Hori} can be viewed~\cite{Hori1975} as an approximate resummation of
the infinite diagrammatic series for the macroscopic conductivity using,
respectively, the self-consistent single-site approximation and the cumulant
expansion.

The averages in Eqs.~\eqref{eqn:Bruggeman} and \eqref{eqn:Hori} are
dominated by typical $\bar{\sigma}$, so it suffices to use $P_n(\bar{n})$
from Eqs.~\eqref{eqn:P_n}. A straightforward numerical solution of these
equations then gives: $\sigma_\text{NP}(\text{Bruggeman}) = 0.48 e^2 / h$ and
$\sigma_\text{NP}(\text{Hori}) = 0.55 e^2 / h$. Taking the difference of the
two as a measure of their accuracy, I arrive at
Eq.~\eqref{eqn:sigma_NP}.

Finally, I briefly comment on experimental implications of the presented theory. For simplicity, I have assumed the dielectric constant of the medium to be the
same on both sides of graphene. It is more realistic
to have graphene at the interface of a half-space with dilelectric
constant $\kappa_1 \sim 1$ and a film of dielectric constant $\kappa_2 \gg 1$
and thickness $D$, whose other side is covered by a metallic gate. In this geometry $\kappa$ is replaced by $(\kappa_1 + \kappa_2) / 2$ while $\mathcal{L}$ becomes
$\min \bigr(\ln ({R} / {\ell})\,,\, \ln({D} / {\ell})\bigl)$.
Small $\alpha$ can be achieved experimentally using ice~\cite{Jang2008tte}, ethanol, and other dielectrics~\cite{Ponomarenko2009eoh}. However, it is difficult to make
$\mathcal{L}$ larger than $\mathcal{L} \sim 5$, for which
Eq.~\eqref{eqn:sigma_NP} predicts $\sigma_\text{NP} \sim 2.5 e^2 / h$. This is
two to three times lower than the measured~\cite{Geim2007rg, Tan2007mos, Chen2008cis, Jang2008tte} $\sigma_\text{NP}$. This may indicate that in experiment the charged impurities are either not exactly coplanar with graphene, or are correlated, or are not the only source of disorder. The last possibility is further corroborated by a very modest increase in mobility~\cite{Ponomarenko2009eoh} away from the NP upon a large \textit{in situ} increase of $\kappa$.


This work is supported by the NSF Grant DMR-0706654. I am grateful to L.~Levitov and B.~Shklovskii for comments on the manuscript.

\vspace{-0.20in}


\end{document}


\title{Neutrality point of graphene with coplanar charged impurities: Supplementary information}

\author{M. M. Fogler}

\affiliation{Department of Physics, University of California San Diego, La
Jolla, 9500 Gilman Drive, California 92093}

\date{\today}

\begin{abstract}

This note contains technical comments on some of the equations presented in the main paper [M.~M.~Fogler, arXiv:0810.1755; Phys. Rev. Lett. \textbf{103}, 236801 (2009)].

A copy of this note is also available as EPAPS Document
No.~E-PRLTAO-103-110952 at the AIP server \url{http://www.aip.org/pubservs/epaps.html}.

\end{abstract}


\maketitle

\noindent$\blacktriangleright$ Equation \eqref{eqn:g_near}.
This equation can be written as
\begin{equation}\label{eqn:g_near}\tag{12}
\varepsilon = \frac{2 \sqrt{\pi}}{3}\, \hbar v |\bar{n}|^{3/2}
                 + \delta \varepsilon\,,
\quad
\delta \varepsilon = -\frac{3 \varepsilon}{4 |\bar{n}| \ell^2}
 \ln \left(\frac{k_F}{\Lambda} \right).
\end{equation}
The first term is the energy density of the state with uniform density $\bar{n}$. All interaction effects are assumed to be included in the renormalized Fermi velocity $v$. In principle, this makes $v$ a function of $\bar{n}$. However, this function is very slow~\cite{Gonzalez2001eei}, and so we treat it as a constant.
The correction term $\delta \varepsilon$ arises due to interaction of electrons with large-$k$ harmonics $N_{\mathbf{k}}$ of the impurity charge density. The corresponding energy density can be expressed in terms of the linear-response dielectric function $\epsilon(k)$ of graphene:
\begin{equation}\label{eqn:delta_varepsilon}
\delta \varepsilon = \frac{1}{2 A^2} \sum\limits_{k > \Lambda}
\left[
\frac{U_\mathbf{k}}{\epsilon(k)} - \frac{U_\mathbf{k}}{\epsilon(\infty)}
\right] \avr{N_{\mathbf{k}} N_{-\mathbf{k}}}
= \frac{\hbar v}{8 \alpha \ell^2} \int\limits_{\Lambda}^{\infty} d k\,
\left[
\frac{1}{\epsilon(k)} - \frac{1}{\epsilon(\infty)}
\right]\,.
\end{equation}
The subtraction in the integrand ensures convergence. (It amounts to a constant shift of no significance for the problem studied.)
Suppose that $\Lambda \gg 4 \alpha k_F$, then
the dominant contribution to the integral comes from
momenta $\Lambda < k < k_F$, at which $\epsilon(k) \simeq 1 + (4 \alpha k_F / k)$. 
Using this approximation and replacing the upper integration limit by $k_F$, we obtain Eq.~\eqref{eqn:g_near}. A more careful calculation that utilizes the full expression~\cite{Ando2006sea} for $\epsilon(k)$ confirms this result.

\noindent$\blacktriangleright$ Equations (3), (15), and (23).
In the main text I allude to
a ``standard expression'' for the two-point probability distribution function (PDF) of $\bar{\Phi}(\mathbf{r})$. This expression is given by~\cite{Isichenko1992pst, Fogler2004nsa}:
\begin{equation} \label{eqn:P_PhiPhi}
P_{\Phi \Phi} (r, f, f^\prime\,) = \frac{1}{2 \pi \sqrt{Q(r)}}
                \exp \left[
\frac{\bar{K}(r)}{Q(r)} f f^\prime - \frac{\bar{K}(0)}{2 Q(r)} \left(f^2 + f^{\prime}\, ^2\right)
                \right]\,,
\quad
Q(r) \equiv \bar{K}^2(r) - \bar{K}^2(0)\,.
\end{equation}
It is easy to see that it respects the identity
$\avr{\bar{\Phi}(0) \bar{\Phi}(\mathbf{r}) } \equiv \bar{K}(r)$ and
gives the correct PDF of a local $\bar{\Phi}(\mathbf{r})$:
\begin{equation} \label{eqn:P_Phi}
P_\Phi(f)  = \int d f^\prime P_{\Phi \Phi}(r, f, f^\prime\,)
           = \frac{1}{\sqrt{2 \pi \raisebox{0pt}[2.00ex]{$\bar{K}(0)$}}}
                          \exp \left(
           -\frac{f^2}{2 \raisebox{0pt}[2.20ex]{$\bar{K}(0)$}}
                               \right)\,.
\end{equation}
On the other hand, if the variance of $\bar{\Phi}$ is computed using the constrained PDF $P_{\Phi \Phi}(r, f, f^\prime = 0)$, Eq.~(23) is obtained.

Next, substituting Eq.~\eqref{eqn:P_PhiPhi} into Eq.~(15) of the main text and directly evaluating the double integral, I get Eq.~(3).
Finally, to derive the relation between $S(r)$ and $K(r)$ at large $r$, I use the expansion
\begin{equation} \label{eqn:P_PhiPhi_far}
P_{\Phi \Phi}(r, f, f^\prime\,) - P_{\Phi}(f) P_{\Phi}(f^\prime\,) \simeq
            \bar{K}(r) \,
            \frac{d P_{\Phi}(f)}{d f} \,
            \frac{d  P_{\Phi}(f^\prime\,)}{d f^\prime}\,,
            \quad r \gg R\,.
\end{equation}
This formula is rather general and is not limited to Gaussian random fields. Indeed, it can be derived without relying on Eq.~\eqref{eqn:P_PhiPhi} but using the primary definition of the two-point PDF: $P_{\Phi \Phi} \equiv \big\langle \delta \big(f - \bar{\Phi}(0)\big)\, \delta \big(f^\prime - \bar{\Phi}(\mathbf{r})\big)$. The idea is that the statistical dependence of random variables $\bar{\Phi}(0)$ and $\bar{\Phi}(\mathbf{r})$ is due to Fourier harmonics $\Phi_\mathbf{k}$ with $k r \lesssim 1$. At large $r$ such $\Phi_\mathbf{k}$ have a small phase space and can be considered a perturbation to a local $\bar{\Phi}$. Therefore, one can expand the arguments of the $\delta$-functions to the second order in such $\Phi_\mathbf{k}$ and average the result to get Eq.~\eqref{eqn:P_PhiPhi_far}.

Substituting Eq.~\eqref{eqn:P_PhiPhi_far} into Eq.~(15) and integrating by parts twice, I get
\begin{equation}
S(r) \simeq {K}(r) \int d f d f^\prime\, 
            P_{\Phi}(f) P_{\Phi}(f^\prime\,)
            \frac{d \bar{n}(f)}{d f} \,
            \frac{d  \bar{n}(f^\prime\,)}{d f^\prime} = \avr{\chi}^2 {K}(r)\,,
\label{eqn:S_far}
\end{equation}
from which the key result $\chi_0 = \avr{\chi}$ follows. Using Eqs.~\eqref{eqn:S_far} and (14) on can show that $\chi_0$ does not depend on the arbitrary cutoff $\Lambda$ and is given by Eq.~(2) of the main text.

\noindent$\blacktriangleright$ Equation~\eqref{eqn:P_n}.
This equation reads
\begin{equation} \label{eqn:P_n}\tag{16}
P_n(\bar{n})
= \frac{\ell}{2 \sqrt{\pi \mathcal{L} |\bar{n}|}}
\,
 \exp\left(- \frac{|\bar{n}| \ell^2}{\mathcal{L}} \right)
\,, \quad |\bar{n}| \ell^2 \gg 1
\,.
\end{equation}
It directly follows from Eq.~\eqref{eqn:P_Phi} and the
renormalized Thomas-Fermi approximation (RTFA), Eq.~(13), if the second (correction) term therein is neglected. Here I wish to discuss how Eq.~\eqref{eqn:P_n} may be modified at $|\bar{n}| \lesssim 1 / \ell^2$ where the RTFA fails. According to
Eq.~\eqref{eqn:P_n}, such $\bar{n}$ are much smaller than the typical $\bar{n} \sim \mathcal{L} / \ell^2$. They are usually found near the $p$-$n$ junctions (PNJ), i.e., near the contours $\bar{\Phi}(\mathbf{r}) = 0$. This is because by symmetry $\nabla \bar{n}$ and $\bar{n}$ are independent random variables, and so $|\nabla \bar{n}|$ is generally not small when $\bar{n}$ is. Therefore, for small $\bar{n}$ the distance $x = \bar{n} / |\nabla \bar{n}|$ to the nearest PNJ is also small.
Equation~(23) implies that in this case
$
\bar{\Phi}(x) = ({\hbar v} / {\ell})\,
             \sqrt{{\pi} \ln \left( x \Lambda \right)}
             \,\,\xi
$,
where $\xi$ is a Gaussian random variable with zero mean and unit variance. The RTFA, which is a local theory, fails for $x$ smaller than the Fermi wavelength $1 / k_F(x) \sim \hbar v /
\left| \bar{\Phi}(x) \right|$. This failure occurs at
$x < x_{\xi} \sim {\ell} \,/\, {|\xi|}$,
assuming of course that $x_{\xi} \ll R$, i.e., that $\xi$, which is specific for
each PNJ, is not too small. Based on our previous study of graphene
PNJ~\cite{Zhang2008nsa, Fogler2008edg}, we expect that at $x < x_{\xi}$, where
$|\bar{n}| < |\bar{n}(x_\xi)| \sim \xi^2 / \ell^2$, the density profile is
linear: $|\bar{n}(x)| \sim x / x_{\xi}^3$. The local inverse compressibility in
the same domain is given by $\chi^{-1} \sim \hbar v x_{\xi}$. The corresponding $P_n(\bar{n})$ can therefore be written as
\begin{equation} \label{eqn:P_n_PNJ}
P_n(\bar{n}) \simeq P_\Phi(0) \langle \langle \chi^{-1}\,
                         \rangle \rangle
        \sim \ell^2 \mathcal{L}^{-1/2}\,
             \langle \langle \, 1 / |\xi| \, \rangle \rangle
        \sim \frac{\ell^2}{\sqrt{\mathcal{L}}}\,
             \ln \frac{1}{|\bar{n}| \ell^2}\,,
\quad
\frac{1}{R^2} \ll |\bar{n}| \ll \frac{1}{\ell^2}\,.
\end{equation}
Here $\langle \langle \ldots \rangle \rangle$ stands for the averaging over
$\xi$ under the indicated above condition $\xi^2 > |\bar{n}| \ell^2$.
Equations~\eqref{eqn:P_n} and \eqref{eqn:P_n_PNJ} match at their common border
of validity $|\bar{n}| \ell^2 \sim 1$. At $|\bar{n}| \ll 1 / R^2$, the
divergence of $P_n(\bar{n})$ presumably saturates.

\noindent$\blacktriangleright$ Equation (19).
This equation states that the local conductivity is determined by the mean-free path $\bar{l}$ computed using a coarse-grained density $\bar{n}$, which is, in turn, obtained by averaging local density $n(\mathbf{r})$ over a neighborhood of a certain size $\Lambda^{-1}$. Here I show that the proper $\Lambda^{-1}$ is of the order of the typical mean-free path $\ell \sqrt{\mathcal{L}}$. To this end consider a region of the sample away from the PNJ where $n(\mathbf{r})$ is large enough and the semiclassical kinetic equation
$
\mathbf{v} \nabla f(\phi, \mathbf{r})
  = \text{St}[f] + \cos \phi
$ applies.
Here $f$ is the deviation of the quasiparticle distribution function from the
equilibrium, $\phi$ is the angle between the directions of quasiparticle
velocity $\mathbf{v}$ and the electric field, and $\text{St}[f]$ is the collision
term.  The importance of density fluctuations can be estimated using a simplifying assumption of isotropic scattering, $\text{St}[f] = -l^{-1}(\mathbf{r}) f$, and then treating $\delta l^{-1}(\mathbf{r}) \equiv l^{-1}\big(n(\mathbf{r})\big) - l^{-1}(\bar{n})$ as a perturbation. Applying the method of characteristics, I get the leading-order result
\begin{equation} \label{eqn:l_eik}
\bar{l} = \int d \mathbf{r}\, \mathcal{G}(r) 
\avr{ \exp \left[ -\int_0^1 \delta l^{-1}(\lambda \mathbf{r}) r d \lambda
           \right]}\,,
\quad
l^{-1}(\mathbf{r})
 = \frac{\pi^{3/2}}{4 \ell^2 |\bar{n}(\mathbf{r})|^{1/2}}\,,
\quad
\mathcal{G}(r) = \frac{1}{2 \pi r}\, e^{-r / \bar{l}}\,.
\end{equation}
Expanded to the the order $\mathcal{O}(\delta l^{-1})$, it reads
$
\bar{l} = {l}(\bar{n})
                 + {l}^2(\bar{n}) \int d \mathbf{r}\, \mathcal{G}(r) 
            \langle \delta l^{-1}(0) \delta l^{-1}(\mathbf{r}) \rangle
$,
which can also be derived diagrammatically. It is easy to see then that the difference between $\bar{l}$ and ${l}(\bar{n})$ is small when the relative change in $\bar{n}(\mathbf{r})$ over the distance of ${l}(\bar{n})$ is much smaller than unity. In turn, this is satisfied when the same can be said about the potential $\bar{\Phi}(\mathbf{r})$. On the other hand, Eqs.~\eqref{eqn:P_PhiPhi} and (13) entail
\begin{equation}
\left|
\frac{\bar{\Phi}(\mathbf{r}) - \bar{\Phi}(0)}{\bar{\Phi}(\mathbf{r})}
\right|
\sim
\frac{1}{|\bar{\Phi}(\mathbf{r})|}\,
\sqrt{\frac{Q(r)}{\raisebox{-1pt}[2.00ex]{$\bar{K}(0)$}} }
\simeq \sqrt{\frac{\ln (r / \ell)}{|\bar{n}| \ell^2}
}\,\,.
\label{eqn:delta_Phi_over_Phi}
\end{equation}
For $r = l(\bar{n})$ [Eq.~\eqref{eqn:l_eik}] the right-hand side of this expression is indeed much smaller than unity as long as $|\bar{n}| \ell^2 \gg 1$, which is precisely the domain of validity of Eq.~(19) indicated in the main text.
